\documentclass[12pt]{amsart}
\usepackage{amsmath, amsthm, amscd, amsfonts,  amssymb ,longtable}

\newtheorem{theorem}{Theorem}[section]
\newtheorem{lemma}[theorem]{Lemma}

\theoremstyle{definition}
\newtheorem{definition}[theorem]{Definition}
\newtheorem{example}[theorem]{Example}

\newtheorem{remark}[theorem]{Remark}
\numberwithin{equation}{section}

\begin{document}

\title[On distance function among finite set of points]{On distance function among finite set of points }

\author[H.Ghahremani Gol, A.Razavi,F.Didehvar]{HAJAR  GHAHREMANI  GOL,$^1$ ASADOLLAH RAZAVI ,$^2$ Farzad didehvar,$^3$ }

\address{$^{1,2,3}$ Faculty of Mathematics and Computer Science, Amirkabir University of Technology, 424 Hafez Avenue, 15914 Tehran, Iran}
\email{ghahremanigol@aut.ac.ir}
\email{arazavi@aut.ac.ir}
\email{didehvar@aut.ac.ir}

\subjclass[2000]{Primary 97 P  XX; Secondary 68 R XX.}

\begin{abstract}
In practical purposes for some geometrical problems in computer science
 we have as information the coordinates of  some finite points in surface instead of the whole body of a surface.
  The problem
arised here is:"How to  define a distance function in a finite space?"as we will show  the appropriate function for this purpose is not a metric function. Here we try to define this distance function in order to apply it in further proposes, specially in the field setting of transportation theory and vehicle routing problem.More precisely in this paper we consider VRP problem for two dimensional manifolds in $\mathbb{R}^{3}$.
\end{abstract}\maketitle.
\section{Introduction and Preliminaries }
There are many algorithms for solving problems such as $TSP , VRP$.All these algorithms solve these problems in the case that surfaces are flat Euclidean, but in general these problems must be proposed in two dimensional manifolds.  Here we try to consider the generalized situation.
In practical cases there are finite points in three dimensional space  $\mathbb{R}^3$.
The strategy is to consider a two dimensional surface based on these points, Later we define  function  for the distance between these points in the defined surface.
To do this first we define a  convenient two dimensional manifold respect to these points, and in the second step we define the distance between two points respect to this manifold.

 For the first level least square method have been generalized to define convenient two-dimensional manifold.
 For the second level,it is expected to define distance between two points by finding vertical feet and then geodesic which connect these vertical feet.Although  on the surface this definition seems a natural definition, there are examples which show this definition is not  an appropriate one.
 Therefore we try to define a new suitable definition. In the new definition for every two points $A ,B$ we consider a two- dimensional manifold which passes through given two points  and also it satisfies the least square method for the rest of points, by applying this we define the distance between $A , B$. We consider the length of  the geodesic which is passed through two points $A ,B$ on the derived surface. Further, we show this definition is better than the former one.
In the first algorithm by considering the natural number as fixed, we passed through these points a polynomial(manifold) degree $n$, like
\begin{equation}\label{polynomial surface}
    z=\sum_{d=0}^{n-i}\sum_{i=0}^{n}a_{i,(n-d-i)} x^{i}y^{(n-d-i)}
\end{equation}

 In the second algorithm to find the distance between two points $A, B$ as we have seen in the first algorithm,
  The manifold should met the following requirements:
  \\(1) The manifold passes through these two points.
  \\(2) This manifold is chosen such that it satisfies the least square conditions for the rest of points.
 \\Let we consider the following example in order to show that why we prefer the second definition.
\begin{example}
 Consider a finite set of points $\{P_1 , P_2,...,P_n\}$ in $\mathbb{R}^3$ such that $\{P_1 , P_2 , ...,P_{n-1}\}$ is encompassed by a circle $C(0 ,r)$
in the $xy$ -plane and $P_n$ is $P_n=(0 , 0, d)$
with $d\gg r$.
 Regarding the first definition distance between $P_1$ and $P_n$ is less than $r$(for large enough $n$ ), but in accordance with the second definition distance between $P_1$ and $P_n$ is close to $d$ as we expect intuitively.
 \end{example}
The second definition is not a definition of metric as we see later.
In the next section,we describes process to fined the distance between each pair of points in the finite set of points.
\section{generalized least square method}
\label{required}
 There  exist numerical methods for obtaining  a useful two dimensional manifold which  can be appropriate for fitting points $P_1 ,P_2,...,P_k$ in $\mathbb{R}^{3}$. We would like to find a surface of degree $n$ can be fitted by  $P_1 ,P_2,...,P_k $.At the first step we write complete form of equation of  degree $n$ as follows:
 $$z=\sum_{d=0}^{n-i}\sum_{i=0}^{n}a_{i,(n-d-i)} x^{i}y^{(n-d-i)}$$
At the second step  we consider the sum of square differences of this equation with the given points , namely:
$$D=\sum_{k=1}^{m}\bigg(\bigg[\sum_{d=0}^{n-i}\sum_{i=0}^{n}a_{i,(n-d-i)} x_{k}^{i}y_{k}^{(n-d-i)}\bigg]-z_{k}\bigg)^{2}$$
At the third step we find parameters such that this sum would be minimized.Then by derivating with  respect to these parameters,
and equaling it to zero we obtain the following equations.
$$\{\frac{\partial D}{\partial a_{i,(n-d-i)} }=0    ~~~~~~  for ~~~~~~~~~~ i=0,..,n   ;  d=0,...n-i.\}$$
By solving this system of equations we have a polynomial manifold.
This is an inherent generalization of least square method to find a line passing a set of points.
It is worth mentioning to note that
in solving this system there is a matrix with the possibility of having the determinant equal to zero,in the case that the determinant of matrix is equal to zero we have infinite surfaces that can be fitted with $P_1 ,P_2,...,P_k$.
\begin{example}
 For points $P_1 ,P_2,...,P_{10} $ such that $P_i=(x_i , y_i , z_i)$ , finding a surface using generalized least square method leads to a matrix with determinant equal to zero. Therefore we have infinite classes of parameters to define a surface.\\
 \\

\begin{center}{\bf Table 1. The Coordinate of Points}\end{center}~

\begin{center}
\begin{tabular}{|c|c|c|c|c|c|c|c|c|c|c|}
  \hline
  $p_i$ & $p_1$ & $p_2$ & $p_3$ & $p_4$ & $p_5$ & $p_6$ & $p_7$ & $p_8$ & $p_9$ & $p_{10}$  \\
  \hline
  $x_i$ & 3 & -3 & -1 & -2 & 0 & 34 & 12 & 5 & 2& 10 \\
  $y_i$ & 0& 0 & 1 & 1 & 0 & 3 & 9 & 12& 5& -4 \\
  $z_i$ & 0 & 7 & 6 & 2 & 4 & 12 & 12 & 18& -3 & -5 \\
  \hline
\end{tabular}
\end{center}~


\end{example}

\section{Distance function}
\begin{definition}
Let $X$ be a finite set of points in $\mathbb{R}^3$, for any $n\in \mathbb{N}$, distance function $d_n:X \times X \longrightarrow\mathbb{R}$   has been defined as follows:\\
We consider $x,y \in X$ ,$d_n(x,y)$ is the length of  geodesic which has passed through $\{x,y\}$ on the surface of degree $n$ , and satisfies  the condition of least square for the rest of points.
\end{definition}

At first it must be showed that this function is well defined.i.e,for any two points $x,y$   there is a  unique surface which passes through points and  satisfies least square condition for the other points, Uniqueness of  the surface is related to the  determinant of the equation of the system  for least square.
i.e,if determinant is non-zero then  related system has a unique solution for coefficients of surface,therefore we have a unique surface with desired conditions.
But if determinant is zero,
there are infinite surface that satisfy conditions. We show in the next theorems we are able to define the distance function as desired.

 According to this definition, $2$
dimensional manifold of $n$ degree which has passed two points $P_1 ,P_2$ should be obtained at first (we want to obtain length of this points.)
and with respect to other points satisfies the least square conditions.
The type of equations will  change in comparisons with the mentioned method.
In the new equations, two primary equations show fitness of  $P_1 ,P_2$ on the surface,the remaining equations should have Maximum fitness in the
 respect to other points.
Thus  we  have  the following equations :
$$\left\{
  \begin{array}{lll}
   z_{1}=\sum_{d=0}^{n-i}\sum_{i=0}^{n}a_{i,(n-d-i)} x_{1}^{i}y_{1}^{(n-d-i)} & \hbox{}\\
z_{2}=\sum_{d=0}^{n-i}\sum_{i=0}^{n}a_{i,(n-d-i)} x_{2}^{i}y_{2}^{(n-d-i)} & \hbox{}\\
   \frac{\partial D}{\partial a_{i,(n-d-i)} }=0, & \hbox{i=0,..,n  ~~~d=0,...n-i} \\
  \end{array}
\right.$$
Now we compute  the length of geodesic passing through  two points $(P_1 ,P_2)$.
 Let $M_C(P_1,... ,P_k)$ be matrix associated to  system equation $\frac{\partial D}{\partial a_{i,(n-d-i)} }=0$. One possible problem during solving the equation system is the possibility of zero determinant of  $M_C(P_1,... ,P_k)$.
\begin{theorem}
 Consider a set of points $X=\{P_0,P_1,... ,P_k\}$,assume $M_C(P_0,P_1 ,
\\P_2 , ...,P_n)$
was equation system of least square method. For any $P_0 \in X$ there is sequence of points $P^{'}_{0,1} , P^{'}_{0,2} , P^{'}_{0,3} , ...$
such that $\lim_{i \rightarrow \infty}P^{'}_{0,i}=P_{0}$ and for any $k \in N$ ,determinant of $Y_{k}=\{P^{'}_{0, k} , P_1 , P_2 , ...,P_n\}$ was non-zero.
\end{theorem}

 \begin{theorem}

 If $C$ is a surface such that $\lim_{i \rightarrow \infty}C_{i}=C$, then the limit of geodesics $g_i$ which passed through between to points $P_n ,P_m$ is $g$,i.e
$\lim_{i \rightarrow \infty}g_{i}=g$

 \end{theorem}

\begin{remark}

 $C_i$ have been obtained by least square methods and $C_i$  are polynomial  surfaces (\ref{polynomial surface}), condition $\lim_{i \rightarrow \infty}C_{i}=C$ is equivalent to $\lim_{i \rightarrow \infty}[a_{k,l}]_{i}=c_{k,l}$
\end{remark}


Let $C$ be a manifold and the determinant of related matrix to $C$ has zero determinant i.e $det(M_C)=0$. Now we consider $\{C_i\}_{i=1}^{\infty}$  such that all of these surfaces passed through $x,y$  and the determinant of the associated matrix of them is not zero. They define the sequence of geodesics $\{g_i\}_{i=1}^{\infty}$. By previous Theorems  $\lim_{i \rightarrow \infty}g_{i}=g$ such that $g$ is  the geodesics of $C$ and $\lim_{i \rightarrow \infty}\|g_{i}\|=\|g\|$. So distance function is well defined.

\begin{definition}
We are capable  to generaliz the previous definition to $\hat{d}_X:\mathbb{R}^3\times\mathbb{R}^3\longrightarrow \mathbb{R}$ such as for any $x , y \in \mathbb{R}^3$ ,we define $\hat{d}$ as follows: $\hat{d}_X(x,y)=d_{X\cup \{x,y\}}(x,y)$.
\end{definition}
we call $(\mathbb{R}^3 , \hat{d})$ 3-finite distance space.
In a natural way we are able to generalized the above theorem to $\mathbb{R}^n$.
we call $(\mathbb{R}^n , \hat{d})$   n-finite distance space.
In this paper we succeed to define the distance function to find distance between points in two dimensional manifolds. Later on,  we will  apply it to some problems in transportation theory.

\begin{lemma}
n-finite distance spaces are not metric spaces.

\end{lemma}

The following counter example shows that $\hat{d}$ is not a metric.
\begin{example}
Let $\{P_1 , ...,P_{10}\}$ be points as show in Table 1.For points $P_1,P_3, P_{10}$ 
\\$d(10 ,1)= 18.84342860 ,d(10 ,3)=12.78704132 ,d(1 ,3)=4.123340349$
\end{example}

\end{document}